\documentclass[aps,pre,showpacs]{revtex4}

\def\la{\langle}
\def\ra{\rangle}
\usepackage[dvips]{graphicx}
\usepackage{epsfig}
\usepackage{float} 
\usepackage{subfloat}
\usepackage[]{subfigure}
\usepackage{latexsym} 
\usepackage{amsmath,amscd}
\usepackage{times}
\usepackage[dvips]{color}

\font\gros=cmbx10 scaled \magstep0
\font\gross=cmbx12 scaled \magstep0
\begin{document}
\baselineskip 0.45cm

\noindent
{\gross HP-sequence design for lattice proteins - an exact enumeration \\
study on diamond as well as square lattice.}

\vspace*{1.0cm}\noindent
{\gros
S. L. Narasimhan$^1$, A. K. Rajarajan$^1$ and L. Vardharaj$^{2}$
}

\vspace*{0.5cm}\noindent
$^1$Solid State Physics Division, Bhabha Atomic Research Center, Mumbai - 400085, India.
$^2$Department of Computer Science and Engineering, National Institute of Technology, Tiruchirapalli - 620015,
India.

\vspace*{1.0cm}

\noindent {\gross Abstract}: 

We present an exact enumeration algorithm for identifying the {\it native} configuration - a maximally compact 
self avoiding walk configuration that is also the minimum energy configuration for a given set of contact-energy schemes; the process is implicitly sequence-dependent. In particular, we show that the 25-step native configuration 
on a diamond lattice consists of two sheet-like structures and is the same for all the contact-energy schemes,
$\{ (-1,0,0);(-7,-3,0); (-7,-3,-1); (-7,-3,1)\}$; on a square lattice also, the 24-step native configuration 
is independent of the energy schemes considered. However, the designing sequence for the diamond lattice walk 
depends on the energy schemes used whereas that for the square lattice walk does not.
We have calculated the temperature-dependent specific heat for these designed sequences and the four energy schemes 
using the exact density of states. These data show that the energy scheme $(-7,-3,-1)$ is preferable to the other
three for both diamond and square lattice because the associated sequences give rise to a sharp low-temperature peak.
We have also presented data for shorter (23-, 21- and 17-step) walks on a diamond lattice to show that this
algorithm helps identify a unique minimum energy configuration by suitably taking care of the ground-state
degeneracy. Interestingly, all these shorter target configurations also show sheet-like secondary structures.

\vspace*{0.5cm}\noindent
{\gross Keywords}:~Lattice proteins, Self Avoiding Walks, HP-design, Specific heat

%\date{\today}

\section*{\leftline{\gross Introduction}}

\noindent {\it Brief review} - Proteins are complex biopolymers that are made up of various sequences of 
amino acids; their biological functions are intimately related to the unique and stable 
conformational structures, referred to as their 'native' states. They are
stable not only with respect to moderate fluctuations in environmental conditions 
but also with respect to minor mutations of their amino acid sequences \cite{Li1}. Their
stability and uniqueness imply that they should correspond to deep funnel-like 
minima of the free energy landscape \cite{Anfinsen}. Since the energy of a conformation depends 
on the amino acid sequence, it is reasonable to expect that nature favors only 
those sequences whose associated energy landscape has a pronounced global minimum.
This suggests an implicit relationship between the conformational state of a protein 
and the amino acid sequence that associates energy with that structure.

There are two aspects to a study of the relationship between structure and sequence -
(a) 'Folding', the process by which a given chain of amino acids assumes a specific 
native conformation of a protein \cite{Dill1} and (b) 'Design', the choice of an amino acid sequence 
for which a given conformation will be the native state of a protein \cite{Shakh1}. 

The number of native structures that correspond to the functional proteins is known to be much 
less, at least by two orders of magnitude, than the total number of proteins \cite{Govind}. This 
implies a many-to-one mapping of amino acid sequences onto the set of all native 
structures. Given the fact that an arbitrarily chosen sequence of amino acids 
cannot represent a protein, identifying the unique set of representative sequences 
that correspond to a given native structure is a challenging problem for the following 
reason.

Since there are $20$ different types of amino acids, the total number of possible 
sequences for a chain consisting of $N$ amino acid residues is $20^N$; also, there 
are a large number of conformations, other than the native structure, available for 
a given chain of amino acid residues. Searching for a funnel-like global free energy 
minimum in the combined sequence-structure space is a formidable task. This necessitates 
building simplified models of protein that can throw light on some basic aspects of
sequence-structure relationship.

The observation that the hydrophobic residues are normally sequestered to the interior 
of a protein's native structure implies the dominant role of hydrophobic effects
in the folding process; this, in turn, suggests a grouping of the twenty naturally 
occurring amino acids into two broad classes - hydrophobic (H) and polar (P) \cite{Go}. 
This enormously reduces the size of the sequence space to be explored. 

By further restricting the dihedral angle to assume only a certain number of 
discrete values in the range $0$ to $2\pi$, we ensure that the number of possible 
conformations of a finite chain of amino acids is denumerably finite.  An off-shoot 
of such coarse-graining schemes is a lattice-protein model \cite{Dill2} in which Self Avoiding 
Walks (SAW \cite{SAW}) on a lattice are taken to represent the various possible conformations 
of an amino acid chain while a two lettered HP-sequence represents the sequence of 
residues.

Within the framework of such a lattice-protein model, it is possible to search for 
a unique set of HP-sequences that correspond to a given SAW configuration, referred to 
as the 'target' configuration. The choice of the target configuration depends on the 
actual systems to be modeled and studied; for example, If sequence-design 
for compact, globular proteins is the problem of interest, then a lattice
SAW having a maximum number of non-bonded nearest neighbor pairs, called 'contacts' 
could be chosen as the target configuration \cite{Shakh2}. The designability of a target
configuration depends on the nature of interactions between monomers in contact \cite{Pande}.

Given a scheme for associating energy with a contact, it will be of general interest 
to devise robust algorithmic rules for identifying a unique target configuration along 
with its designing sequence. Lot of work reported in the literature focus on designing 
HP-sequence for SAW configurations within regular lattice shapes such as the $6\times 6$ square 
or the $3\times 3\times 3$ cube, and also within different shapes on a square lattice \cite{Peto}. 

Different secondary structural elements are known to be favored by different lattices \cite{Godzik}; 
for example, at low resolution (3-4 A rms), a cubic lattice favors $\alpha$-helix structure 
whereas a diamond lattice favors $\beta$-sheet structure. On the basis of a detailed comparative 
study of various lattices, Godzik et al. \cite{Godzik} have shown that SAWs on lattices with higher 
coordination number provide better structural representation of a protein, even though 
lattice effects are minimal as far as folding pathways are concerned.

\noindent {\it Motivation} - Since the various contact-energy schemes considered in the literature are 
based on statistical estimates 
of the average effective inter-residue interactions of real proteins \cite{Miyazawa}, it is of interest 
to study how sensitive is the choice of a target configuration to changes in the contact-energy
scheme. As this depends also on the specific sequence used, choosing a target configuration itself should
be a sequence-dependent process. In fact, the many-to-one mapping of amino acid sequences onto the set of 
all native structures implies that a given native structure should not be very sensitive to the set of
contact-energy schemes underlying the set of sequences designing it.
There is also a related question of realizing a funnel-shaped energy-
landscape which, in a milder form, implies reducing the diversity of target configurations as much as
possible. Subsequently, for a target configuration chosen, sequence-design is a process of finding a sequence 
that gives maximum depth to the funnel. It is worthwhile investigating whether a mutually consistent 
treatment can be given to these two processes - namely, (i) identification of a target configuration which 
is not very sensitive to changes in the contact-energy scheme and (ii) designing a sequence that ensures 
maximum energy-gap between the target configuration and the others. 

In this paper, we describe an exact enumeration-based search algorithm in the combined 
sequence-structure space for identifying the target (native) SAW configuration as well 
as its designing sequence on a diamond as well as on a square lattice. The motivation is to evolve an algorithmic 
criterion for identifying the best energy scheme from among the set of schemes being considered. We show that 
the target configuration (a 25-step compact SAW on a diamond lattice \cite{Hurst} and a 24-step compact SAW on 
a square lattice) is not sensitive to the different energy schemes considered, 
whereas the designing sequences are. The native configuration on a diamond lattice is seen to 
consist of two 'sheets' (Fig.\ref{fig:DesConf}) indicating the appearance of secondary structural elements even 
for such short chains. 
From the Density of States and the numerical values obtained for the zeros of the partition function, we calculate 
the temperature-dependent specific heat for all the designing sequences and the energy schemes. 
However, {\it ground-state degeneracy} ({\it i.e}., the number of target configurations)
increases for shorter walks.
We have also that the target configurations of shorter (23-, 21- and 17-step) walks on a diamond
lattice show sheet-like secondary structures.

%This paper is 
%organized as follows. After a brief description of the design-methodology in the next section, we present 
%the results of our study of diamond and square lattice walks; we summarize our results and conclude in the 
%last section.

\section*{\leftline{\gross Methods}}

\subsection*{\leftline{A. Self Avoiding Walks (SAW) on a lattice \cite{SAW}.}}

An ordered set of points on a given lattice of coordination number $z$, called a 'chain' or a 'walk', 
can be specified either by listing the coordinates of all the points or by specifying the starting point 
and the sequence of direction-vectors between consecutive points of the walk. It is a SAW if the points 
in the list are all distinct, which is to say that the walk does not visit a site already visited once.

Different lattices have different sets of direction-vectors. For example, a cubic lattice has the
set of six direction-vectors $\{(\pm 1,0,0), (0,\pm 1,0), (0,0,\pm 1)\}$ whereas a diamond lattice has 
two sets of direction-vectors, namely $\pm \{(1,1,1), (-1,-1,1),(1,-1,-1),(-1,1,-1)\}$ , which have to be 
used alternately in going from site 
to site.

%and $\{(-1,-1,-1), (1,1,-1),(-1,1,1),(1,-1,1)\}$

SAWs are known to have certain universal properties, independent of the lattice in which they are
embedded. For example, the total number of $N$-step SAWs on any given lattice has an asymptotic 
scaling form \cite{SAW},
\begin{equation}
\label{eq:TotNumSAW1}
Z_N = a_N N^{\gamma -1}\mu ^N
\end{equation}
where $\gamma$ is a universal exponent whose value depends only the dimensionality of the lattice,
$\mu$ is the effective coordination number \cite{footnote1} and $a_N$ is a lattice-dependent proportionality
constant.The average 'size' of an $N$-step SAW is specified either in terms of the mean squared 
Radius of Gyration, $\la R_g^2(N)\ra$, 
or equivalently in terms of the mean squared end-to-end distance, $\la R_e^2(N)\ra$,  
which are again of the form,
\begin{eqnarray}
\label{eq:RadGyrSAW}
\la R_g(N)\ra = b_N N^{2\nu} \\
\la R_e^2(N)\ra = c_N N^{2\nu}
\end{eqnarray}
where $\nu$ is a universal exponent \cite{footnote1} and, $b_N$ and $c_N$ are lattice dependent proportionality 
constants. In the limit of large $N$, the ratio, $b_N/c_N$, is known to converge to the values,
$0.14$ and $0.16$, in two and three dimensions respectively.

An $N$-step SAW configuration may have sites that are nearest neighbors on the lattice but far away along 
the chain; such 'non-bonded' pairs of nearest neighbors are referred to as {\it contacts}. If $Z_{N,m}$ 
denotes the number of $N$-step SAWs having $m$ contacts, we can also write
\begin{equation}
\label{eq:TotNumSAW2}
Z_N = \sum _{m=0}^M Z_{N,m}
\end{equation}
where $M$ is the maximum number of contacts an $N$-step walk can have. Labeling the sites of a configuration 
in a sequential order leads to a concise list of contacts - namely, $\{ (j,k)\mid j,k = 1, 2, 3, ..., N; j<k \}$ -
called its {\it contact-map}.

\subsection*{\leftline{B. HP-model \cite{Go}.}}

In this model, conformations of proteins are assumed to be represented by lattice SAWs whose sites are
occupied either by hydrophobic (H) or by polar (P) residues. So, an HP-protein is a point,
$\{\cal{C}_N,\vec{\sigma}\}$, in the combined sequence-structure space where $\cal{C}_N$ is the backbone 
SAW configuration to which is assigned the binary sequence 
$\vec{\sigma} = \{\sigma _j \mid \sigma _j = H \mbox{ or } P; j = 1, 2, 3, ..., N\}$. We treat an amino 
acid residue (H or P) in the chain as a structureless bead and refer to it simply as a 'monomer' of type H or P.

Each monomer is connected (or bonded) to two other monomers in the chain and hence 
can make a maximum of $(z-2)$ contacts; however, at either end of the chain, it can have a maximum 
of $(z-1)$ contacts. Interactions between monomers contribute to the energy, $E({\cal C}_N,\vec{\sigma})$,
of an HP-protein. Bond-energy contribution to $E({\cal C}_N,\vec{\sigma})$ can be ignored as it is a constant
independent of the chain-configurations; so, it is only the interactions between non-bonded nearest neighbors in 
the chain that contribute to $E({\cal C}_N,\vec{\sigma})$. Different configurations can have different sets of
contacts and hence can have different values of $E({\cal C}_N,\vec{\sigma})$. In terms of a basic unit of energy,
$\epsilon _0$, we may write
\begin{equation}
\label{eq:ChainEnergy}
E({\cal C}_N,\vec{\sigma}) = \sum _{(j,k>j+1)} C_{jk}U(\sigma _j, \sigma _k)
\end{equation}
where $C_{jk}$ is the contact-map given by
\begin{equation}
C_{jk} = \left\{
                \begin{array}{ll}
                  1 \qquad & \mbox{ for non-bonded nearest neighbors }i ~\mbox{and}~ j\\
                  0 \qquad & \mbox{ othewise}
                \end{array}
         \right.
\label{eq:hgt}
\end{equation}
and $U(\sigma _j,\sigma _k)$ denotes the energy associated with the contact $(j,k)$ whose value depends on the 
type of monomers (H or P) in contact - namely, $U(\sigma _j,\sigma _k) = \{ u_{HH}, u_{HP}, u_{PP}\}$. In other 
words, $U(\sigma _j,\sigma _k)$ depends not only on the configuration through its contact-map but also on the 
HP-sequence considered. 

Based on a statistical analysis \cite{Miyazawa} of the various interactions between the amino acid residues of 
real proteins, the values of $U(\sigma _j,\sigma _k)$ have been shown \cite{Li2} to satisfy the inequalities,
$u_{HH} < u_{HP} < u_{PP} \mbox{ and }(u_{HH} + u_{PP}) < 2u_{HP}$. Together, these inequalities imply the 
dominant role of hydrophobic interaction in getting the H-type of monomers together and also a general tendency 
of the two types of monomers to segregate. The number of two-lettered HP sequences leading to a unique
ground state, for a given scheme of energy values, is taken to define the 'designability' of a model.

The simplest energy scheme is the one in which the only interaction considered is that between the H-type monomers, 
{\it i.e}., $(u_{HH}, u_{HP}, u_{PP}) = (-1, 0, 0)$ \cite{UngerToma}. It is equivalent to saying that the 
minimum energy configurations, referred to as the ground states, are those with maximum number of contacts. 
This model brings out the gross qualitative features that are comparable to those of real proteins; however, 
it leads to highly degenerate ground states with small number of 'designing' sequences ({\it i.e}., sequences 
with unique ground state).

A modified version of this simple scheme is called the shifted HP-model in which the $u$'s are assigned the 
values, $(u_{HH}, u_{HP}, u_{PP}) = (-2, 1, 1)$ \cite{Chan}. Yet another widely studied scheme is defined by the 
set of values, $(u_{HH}, u_{HP}, u_{PP}) = (-2.3, -1, 0)$  \cite{Li2}. It is not necessary that the ground
states in these schemes should correspond to maximally compact ({\it i.e}., maximum number of contacts)
configurations. However, they lead to less degenerate and better designable structures. As pointed out by 
Pande {\it et al}. \cite{Pande}, similar interactions lead to similar degrees of designability while different
interactions lead to different patterns.  

\subsection*{\leftline{C. Target structure.}}

By definition, a {\it target} structure is a SAW configuration, usually a maximally compact one, that is 
assigned minimum energy by a set of HP-sequences within a contact energy scheme, 
say $U (\equiv \{u_{HH}, u_{HP}, u_{PP}\})$. The value of the minimum energy realized depends on the configuration, 
and is given by
\begin{equation}
\label{eq:EgyTarget}
E_0({\cal C}_N,n_H;U) = \mbox{min}\{ E({\cal C}_N,\vec{\sigma}_j)\mid j = 1, 2, 3, ..., M \}; 
                \qquad M = \frac{N!}{n_H! (N-n_H)!}
\end{equation}
where $n_H$ is the number of H-type monomers that are required to be part of the HP-chain. Let 
$b({\cal C}_N,n_H;U)$ denote the the number of HP-sequences assigning minimum energy, 
to the configuration ${\cal C}_N$. Then, it is more designable than another 
configuration, say ${\cal C}'_N$, having the same minimum energy if $b({\cal C}_N,n_H;U)> b({\cal C}'_N,n_H;U)$. 
Given a set of configurations, $\{ {\cal C}_N\}$, 
those having the energy $E_g \equiv \mbox{min} \{ E_0\}$ are the ground state configurations each of which
in turn is characterized by a certain value of $b$. Configurations of interest are those, $\{ {\cal C}_N^g\}$,
that are characterized by the values $(E_g,b_g)$ where $b_g \equiv \mbox{max}\{ b\}$. 

It is not necessary that the set of designing sequences be the same for all the configurations,
${\cal C}_N^g$. Also, the sets of designing sequences as well as the set of designing configurations,
$\{ {\cal C}_N^g\}$, may be different for different contact energy schemes. In order to confirm this explicitly,
we have considered the following contact energy schemes:
\begin{eqnarray}
\label{eq:EgyScheme1}
U_1 & = & (-1, 0, 0) \\
\label{eq:EgyScheme2}
U_2 & = & (-7, -3, 0) \\
\label{eq:EgyScheme3}
U_3 & = & (-7, -3, -1) \\
\label{eq:EgyScheme4}
U_4 & = & (-7, -3, 1)
\end{eqnarray}
$U_1$ is the standard energy scheme mentioned earlier \cite{UngerToma}; $U_2$ is just three times the 
scheme $(-2.3, -1, 0)$ \cite{Li2}, integerized; $U_3$ includes an attractive PP-contact energy, and is not
related to $U_2$ by any numerical operation such as integer addition or multiplication. All these three 
schemes satisfy the inequalities \cite{Li2} mentioned in the last section. The scheme $U_4$ has a repulsive 
PP-contact energy, and violates one of the inequalities namely, $(u_{HH} + u_{PP}) < 2u_{HP}$; this is 
done just to see what difference it makes {\it vis-a-vis} the other schemes.

\subsection*{\leftline{D. Stability.}}

Levinthal's paradox \cite{Levinthal} refers to the observation that a denatured protein folds spontaneously 
to its native state on very short time scales even though a sequential search for its native state would take 
forever to complete. It was then thought that folding proceeds along specific pathways through intermediate
structures. But, Anfinsen's \cite{Anfinsen} experiment demonstrated that folding is independent of pathways,
suggesting thereby a funnel-like free energy landscape \cite{Wolynes1} in which the native state lies at 
the bottom. While the depth of the funnel corresponds to the energy-stabilization of the native state with 
respect to a denatured state, the width of the funnel corresponds to the conformational entropy of the chain.
This picture provides a conceptual framework for a fast folding kinetics. The existence of a large energy gap 
between the native state and any other conformational state is, therefore, considered  as a prerequisite for 
a model protein. 

As described in the previous section, we may have a set of ground state configurations, $\{ {\cal C}_N^g\}$,
all having the minimum energy $E_g$ and designable to the same degree; that is to say, the number of 
HP-sequences associated with each configuration is the same, namely $b_g$. The sequences themselves may be
different for different configurations. 

\bigskip

\noindent 1. {\it Non-degenerate ground state}:

\bigskip

If there is only one ground state configuration, ${\cal C}_N^g$, and the associated set ${\cal S}^g$ of 
HP-sequences (all giving the same energy, by definition, to ${\cal C}_N^g$), it is necessary to check
whether it is at the bottom of a funnel in the energy landscape.

Since ${\cal C}_N^g$ is the unique ground state, all the other configurations are excited states whose
energies will depend upon the HP-sequence fitted to them. For every sequence $s^g \in {\cal S}^g$, an average 
excited state with energy $\la E_g^x(s_g)\ra$ and the corresponding standard deviation $\Delta _g(s_g)$ can
be obtained. The {\it gap}, $(\la E_g^x(s_g)\ra -E_g)$ is a measure of the funnel-depth and $\Delta _g(s_g)$
is a measure of the funnel-width. The ratio of these quantities \cite{Bryngelson}, 
\begin{equation}
\label{GapParameter}
f({\cal C}_N^g,s_g) \equiv \frac{(\la E_g^x(s_g)\ra -E_g)}{\Delta _g(s_g)}; 
\qquad f({\cal C}_N^g) \equiv \mbox{max}\{ f({\cal C}_N^g,s_g)\}
\end{equation}
can be taken to be a single parameter characterizing the funnel. Sequences having the maximum gap-parameter value, 
$f$, may be deemed the best designing sequences.

\bigskip

\noindent 2. {\it Degenerate ground state}:

\bigskip

If there are more than one ground state, then {\it a priori} all are equally probable and so 
the funnel is likely to have so many wells at the bottom (a flat bottom, in the coarse-grained sense). 
With each of the configuration, ${\cal C}_N^g$, we have a 
set of sequences ${\cal S}^g({\cal C}_N^g)$ all giving it the minimum energy $E_g$. If there is no sequence
common to all the sequence-sets ({\it i.e}., $\cap _{{\cal C}_N^g} ~ {\cal S}^g({\cal C}_N^g) = \phi$), then 
we may identify those configurations for which the gap-parameter, $f ( = \mbox{max}\{ f({\cal C}_N^g)\})$,  
is the maximum. On the other hand, if $\cap _{{\cal C}_N^g} ~ {\cal S}^g({\cal C}_N^g) \neq \phi$, 
then this procedure cannot reduce the degeneracy of the ground state; all the degenerate configurations 
have the same set of common designing sequences. 

It is possible that different contact energy schemes give rise to different sets of ground states. Since 
the estimate of a contact energy is statistical in nature \cite{Miyazawa}, it is not unreasonable to 
speculate that there may be ground state configurations that are insensitive to changes in the 
energy scheme. If such is the case found by repeating the above procedure for different energy schemes
(Eq.(\ref{eq:EgyScheme1}-\ref{eq:EgyScheme4}), for example), then degeneracy is reduced even further.
In other words, we try to identify those configurations for which the gap-parameter, 
$f ( = \mbox{max}\{ f({\cal C}_N^g)\})$, has a set of maximum values corresponding to the set of contact-energy
schemes. 

It is also possible that there are configurations designed by sequences that are mutually inverse
(mirror-image) to each other; for example, $HP_4(H_2P)_2HPH_3P_2HP_3H_2$ is inverse to $H_2P_3HP_2H_3PH(PH_2)_2P_4H$.
In that case, we may check whether the configurations can be mapped into one another by a simple permutation
and relabeling of the direction-vectors; if so, they are equivalent configurations. 
If this procedure leads to just one configuration, then we have a unique ground state along with its set of 
designing sequences; otherwise, this procedure reduces the ground-state degeneracy to a maximum extent.

\bigskip

\noindent 3. {\it Specific Heat}:

\bigskip

Given a sequence and an energy scheme, we use the enumeration algorithm to have an exact count of the number of
configurations having an energy, $E$; The integerized forms of the energy schemes chosen, 
(Eq.(\ref{eq:EgyScheme1}-\ref{eq:EgyScheme4})), ensure that $E$ is an integer between zero and the minimum
value $-E_g$. The total number of an N-step walks, also called the {\it microcanonical} Partition Function, is 
then given by
\begin{equation}
\label{eq:PF1}
Z_N(U) = \sum _{E=0}^{E_g} Z_{N,E}(U)
\end{equation}
which is the same as Eq.(\ref{eq:TotNumSAW2}) for $U=U_1$ \cite{footnote4}. The numbers, $\{ Z_{N,E}(U)\}$, are 
also known as the Density of States (DoS). At finite temperatures, $T$, the DoS will be weighted by the 
Boltzmann factor, $e^{E/T}$; since $E$ is an integer, the weighted sum, $Z_N(U,w)$, called the {\it canonical}
Partition function is actually a polynomial of degree $E_g$ in the variable $w (\equiv e^{1/T})$. We can write
\begin{equation}
\label{eq:PF2}
Z_N(U,w) = \prod _{E=0}^{E_g} (w-w_E)
\end{equation}
where $w_E$s are the (complex) roots of the polynomial equation, $Z_N(U,w)=0$. These roots will occur in the form
of complex conjugate pairs because $Z_N(U,w)$ is a real number. The specific heat at any given temperature is 
then given by
\begin{equation}
\label{SpHeat1}
C(U,w) = w[\ln (w)]^2 \frac{\partial}{\partial w}\left( w\frac{\partial \ln Z_N(U,w)}{\partial w}\right)
\end{equation}
In terms of the real, $u$, and the imaginary, $v$, parts of $w$, we have
\begin{equation}
\label{SpHeat2}
C(U,w) = w[\ln (w)]^2 \sum _{j=1}^{E_g} \frac{[(2w-u_j)v_j^2-u_j(w-u_j)^2]}{[(w-u_j)^2+v_j^2]^2}; \qquad w = e^{1/T}
\end{equation}
Fast folding sequences have been observed \cite{Socci} to give rise to two different peaks in the specific heat 
of which the one at lower temperature is expected to be sharp and prominent.

It must be mentioned here that the specific heat, $C(U,T)$, can also be computed directly from the DoS using the 
expression,
\begin{equation}
\label{SpHeat3}
C(U,T) = \frac{1}{T^2}(\la E \ra ^2 - \la E^2 \ra ); \quad 
         \la E \ra \equiv \frac{\sum _{E=0}^{E_g} Z_{N,E}(U)E ~ \mbox{e}^{E/T}}
                               {\sum _{E=0}^{E_g} Z(N,E) ~ \mbox{e}^{E/T}}
\end{equation}
and it will be the same as $C(U,w)$. However, partition function zeros provide additional corroborating evidence
for considering a peak in the specific heat as indicative of a phase transition, especially when we study small systems.

\section*{\leftline{\gross Results}}

\noindent 1. {\it Algorithm}:

\bigskip

Using an exact enumeration module for generating SAWs on any given lattice, we first dump all N-step SAWs 
having maximum number of contacts. For a given energy scheme, $U$, and a binary sequence having $n_H$ ones,  
it is the set of contacts that determines the energy of a walk; so, we identify and collect all the walks 
with distinct contact-maps.

In order to identify the target configurations ({\it i.e}., walks, to which maximum number of 
HP-sequences assign minimum energy), we generate binary sequences having a specified number of ones ('H's)
and find the energy they assign to the maximally compact walks, taken one by one, by using their contact-maps. 
This way, we have a count of binary sequences that assign minimum energy attainable for each walk and 
a list of those walks having global minimum energy; from this list, we choose walks with which maximum
number of (designing) sequences are associated. These are the (designable) target configurations for a given 
energy scheme, say $\{ {\cal C}_{N,n_H}(g)\}_U$.

By repeating this procedure for all the specified energy schemes, we have sets of designable configurations 
and their associated sets of designing sequences. If there are configurations common to all these sets, say
$\{ {\cal C}_{N,n_H}(g)\} \equiv \cap _U \{ {\cal C}_{N,n_H}(g)\}_U$, we consider only them for further analysis; 
otherwise, we have to consider all these sets one by one. As described in the last section, we use the 
gap-parameter,$f$, for reducing the degeneracy of the ground state as far as possible: 

From the set of sequences designing a given configuration, 
say $ {\cal C}_{N,n_H}^{(j)}(g) \in \{ {\cal C}_{N,n_H}(g)\}$, we identify the ones with maximum values of $f(U)$ for
the energy-schemes; it is possible that the same sequence has maximum value of for more than one energy-scheme.
We do it for every configuration in the set; the set of sequences and their corresponding $f$ values may be different
for different configurations. For example, if there are $K$ configurations under consideration ($j = 1, 2, \cdots K$)
and four energy-schemes, then for each $j$, we have four sequences (same or different) and their
corresponding $f$ values, say $\vec{f}^{(j)} \equiv \{ f^{(j)}_1, f^{(j)}_2, f^{(j)}_3, f^{(j)}_4\}$.
We choose those configurations and their associated sequences to which correspond the maximum of the set 
$\{ \vec{f}^{(j)}\}$. If there are more than one such configuration, then we have to decide whether they are
equivalent with respect to a permutation and relabeling of the direction-vectors. This way, the ground state
degeneracy can be reduced as much as possible.

\bigskip

\noindent 2. {\it Diamond Lattice walks}:

\bigskip

In the case of diamond lattice, we have two sets of direction-vectors, 
namely $\{ v_I(j)\mid j=1,2,3,4\} \equiv \{(1,1,1),(-1,-1,1),(1,-1,-1),(-1,1,-1)\}$ and 
$\{ v_{II}(j)\} = \{ -v_I(j)\}$ which have to be used alternately in going from site to site $-$   
{\it i.e}., if the $K^{th}$ step is taken in the direction $v_I(j)$, then the next step will have to be taken
in one of the three directions $v_{II}(l), l\neq j$ after ensuring that this step avoids previously visited sites. 
A recursive procedure for adding a step in a systematic manner can enumerate all SAWs up to a specified 
maximum length. 

Also, any given configuration can be coded in terms of the direction-labels such as,
for example, 1234231242313123243212132 which codes a 25-step walk. Using such a code and assuming that the walk
starts from position $\vec{r}_0$(from the origin, for example), the positions of the subsequent sites in 
the walk can be generated by knowing the direction-type (I or II) of the first step. We have fixed the first 
two steps in the directions $v_I(1)$ and $v_{II}(2)$ respectively.

Using such exact enumeration procedure, we have generated 25-step SAWs on a diamond lattice \cite{footnote2}
and found that there are 270 maximally compact walks with 12 contacts each, of which 135 walks have distinct
contact-maps. In TABLE \ref{tab:steps25I}, we have presented the number of walks \cite{footnote3} and their 
average end-to-end 
distance as a function of the number of contacts.
\begin{table}[t]
\begin{tabular}{rrr}\hline 
$m$ \ & $Z_{25}(m)$ \ & $\la r_{25}(m)\ra$ \\  \hline 
0 \ & 18409190935 & \ 15.3801 \\
1 \ & 16564041012 & \ 13.4874 \\
2 \ & 9902194362 & \ 12.0026 \\
3 \ & 4583935502 & \ 10.5940 \\
4 \ & 1791649410 & \ 9.3488 \\
5 \ & 606531194 & \ 8.2332 \\
6 \ & 180088606 & \ 7.2606 \\
7 \ & 44257966 & \ 6.4466 \\
8 \ & 8323860  & \ 5.7982 \\
9 \ & 1068592 & \ 5.2175 \\
10 \ & 67036  & \ 4.9458 \\
11 \ & 872 & \ 5.5048 \\
12 \ & 270 & \ 3.7446 \\ \hline 
\end{tabular} \\
\caption{$Z_{25}(m)$: Number ($\times 12$) of 25-step walks making $m$ contacts;
$\la r_{25}(m)\ra$: average end-to-end distance.}
\label{tab:steps25I}
\end{table}
\begin{table}
\begin{tabular}[t]{cccc}\hline 
$U$ \ & $Z_{25}^g(U)$ \ & $E_g$ \ & $S_g$\\ \hline 
$U_1$ \ & 40 \ & -8 \ & 630 \\
$U_2$ \ & 40 \ & -59 \ & 420 \\
$U_3$ \ & 40 \ & -62 \ & 420 \\
$U_4$ \ & 40 \ & -56 \ & 31824 \\ \hline 
\end{tabular}\\ 
\caption{$U$: Energy scheme (Eq.(\ref{eq:EgyScheme1}-\ref{eq:EgyScheme4});
$Z_{25}^g(U)$: Number of 25-step ground state configurations  corresponding to $U$;
$E_g$: Ground state energy; $S_g$: Number of designing sequences.}
\label{tab:steps25II}
\begin{tabular}[t]{ccc}\hline 
$U$ \ & Sequence \ & f\\  \hline 
$U_1$ \ & $H_3P_2HP_2H_2P_2HP_2H_2P_3H_2(PH)_2$ \ & 4.2126\\
$U_2$ \ & $H_3PH_2P_3H_2PHP_2(HP)_2PHP(PH)_2$ \ & 3.9946\\
$U_3$ \ & $(HP)_2H_2P_3H_2P_4H_3P_2H_2(PH)_2$ \ & 4.1622\\
$U_4$ \ & $H_3PH_2P_3H_2PHP_2(HP)_2PHP(PH)_2$ \ & 3.7132\\ \hline
\end{tabular}\\ 
\caption{f: maximum gap-parameter. Subscript of H or P denotes the number of times the corresponding symbol 
         is repeated.}
\label{tab:steps25III}      
\end{table}

\begin{figure} [h]
\centering
\unitlength1cm
  \begin{minipage}[t]{6.0 cm}
    \subfigure{\hspace{-1.0 cm} \includegraphics[width=1.5\textwidth ]{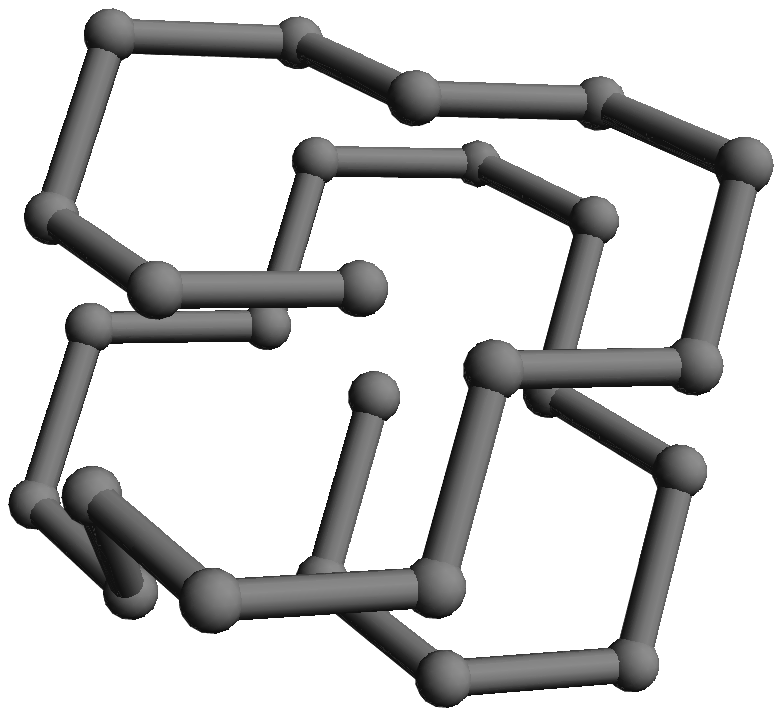}}               
  \end{minipage}
\hfill
  \begin{minipage}[t]{6.0 cm}           
    \subfigure{\hspace{-2.0 cm} \includegraphics[width=1.5\textwidth ]{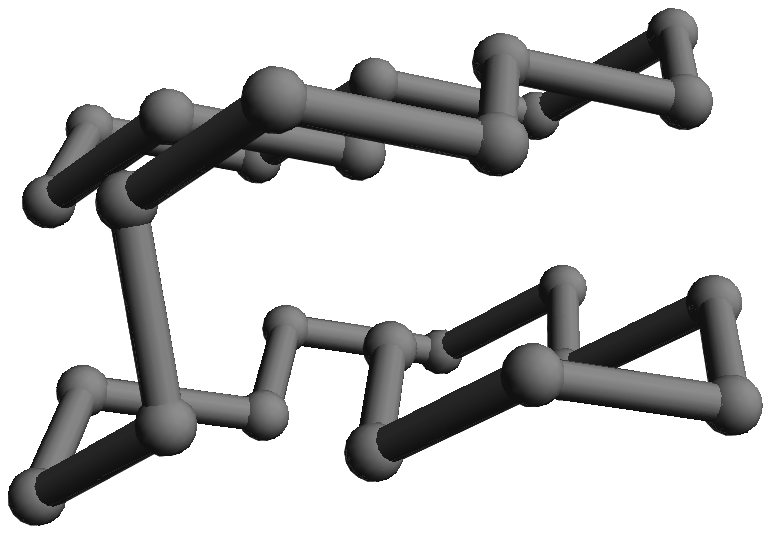}}             
  \end{minipage} 
\vspace{-1.5 cm}          
\caption{Two different views of the most designable configuration, 1231212313124231312323123, that is common 
         to all the four energy schemes considered (Eq.(\ref{eq:EgyScheme1}-\ref{eq:EgyScheme4})) . 
         As is evident from view (b), the structure consists of two 'sheets'.}
\label{fig:DesConf}  
\end{figure}

By fitting all possible 26-bit binary sequences (13 ones and 13 zeros) to each of the 135 maximally compact 
and distinct configurations, we shortlisted designable configurations for each of the energy schemes given in
(Eq.(\ref{eq:EgyScheme1}-\ref{eq:EgyScheme4})) and presented the data in TABLE \ref{tab:steps25II}. For a given 
energy scheme,  
the {\it number} of sequences associated with each of the 40 ground state configurations is the same, and is 
the maximum. The sets of ground states, though 40 in number for all the schemes, are different {\it except}  
for one configuration, 1231212313124231312323123. This is the configuration which is also the ground state
for all the four energy schemes and hence is the most designable ground state configuration. As shown in 
Fig.\ref{fig:DesConf}, this configuration is clearly seen to consist of two sheet-like secondary structures.

Among the sets of sequences designing this configuration, for the energy schemes considered, We found 
420 sequences common to these sets. In other words, these 420 sequences design the configuration irrespective of 
the energy scheme considered and hence may be referred to as the best designing sequences. We used these
sequences and the original set of 135 distinct configurations so as to address the thermodynamic stability of 
the unique ground state configuration obtained (Fig.\ref{fig:DesConf}).

As described above, for every energy scheme ($U_1$ to $U_4$), we find the gap-parameter associated with each
sequence by identifying which of the 135 configurations correspond to 'excited' states. From the list of the
gap-parameters associated with the corresponding sequences, we can choose the sequence with the maximum
gap-parameter. We present the best sequences obtained in TABLE \ref{tab:steps25III}.

We see that the sequences corresponding to $U_2$ and $U_4$ are the same and so, we have three distinct sequences. 
This result suggests
that the basic energy scheme also plays a role in ensuring a large gap between the ground state and the excited 
state configurations. To further confirm that the schemes $U_2$ and $U_4$ lead to the same designing sequence, 
we have rerun the exact enumeration algorithm for computing the Density of States (DoS) for these sequences and 
the energy schemes.

\begin{center}
\begin{figure} [ht]
  \subfigure [] 
     {\includegraphics[width=0.5\textwidth, angle=0]{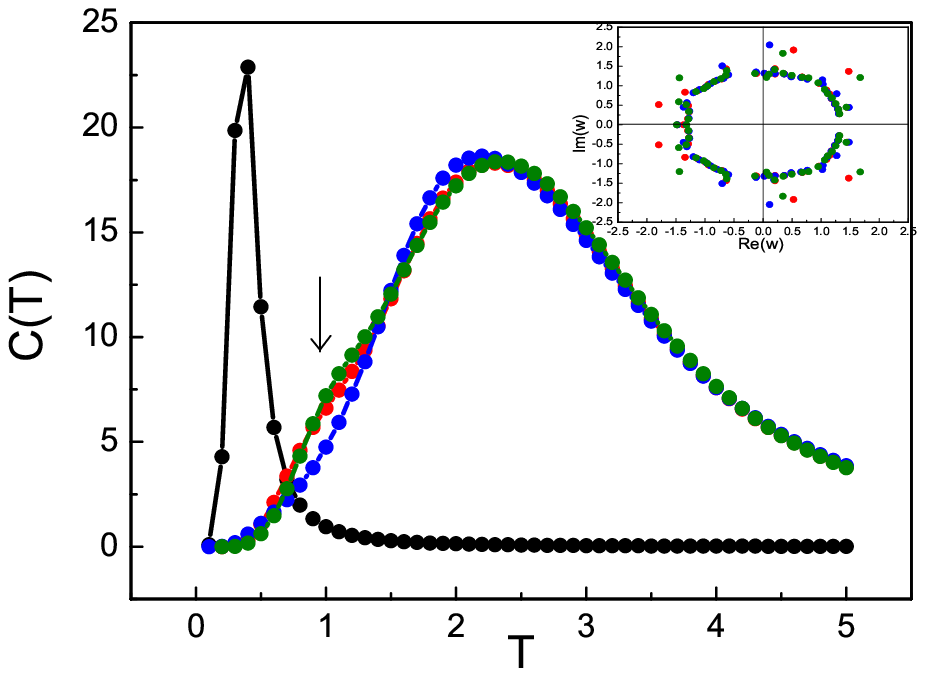}}                        
  \subfigure []
     {\includegraphics[width=0.5\textwidth, angle=0]{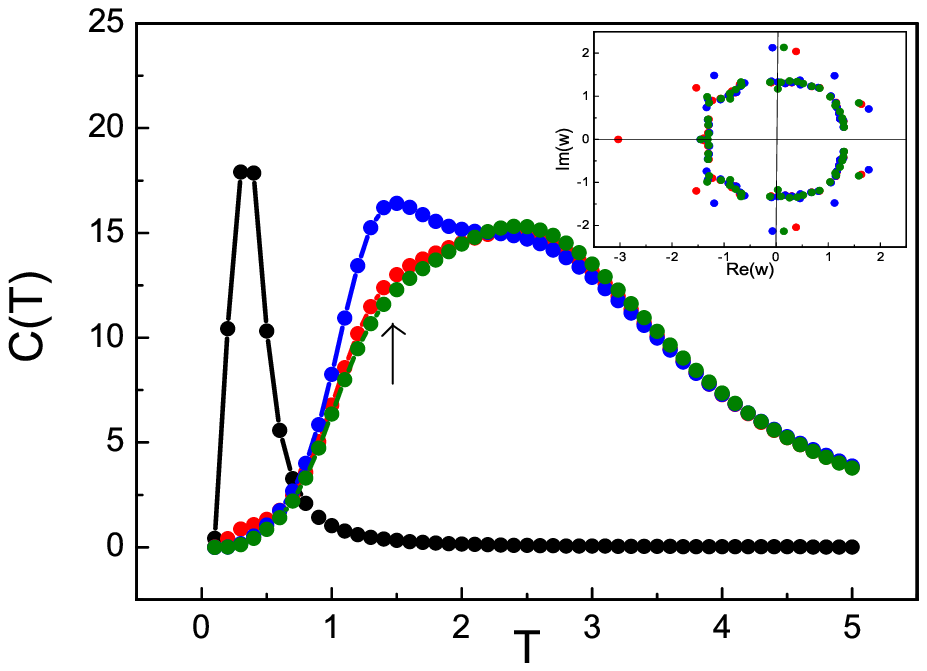}} 
  \subfigure []
     {\includegraphics[width=0.5\textwidth, angle=0]{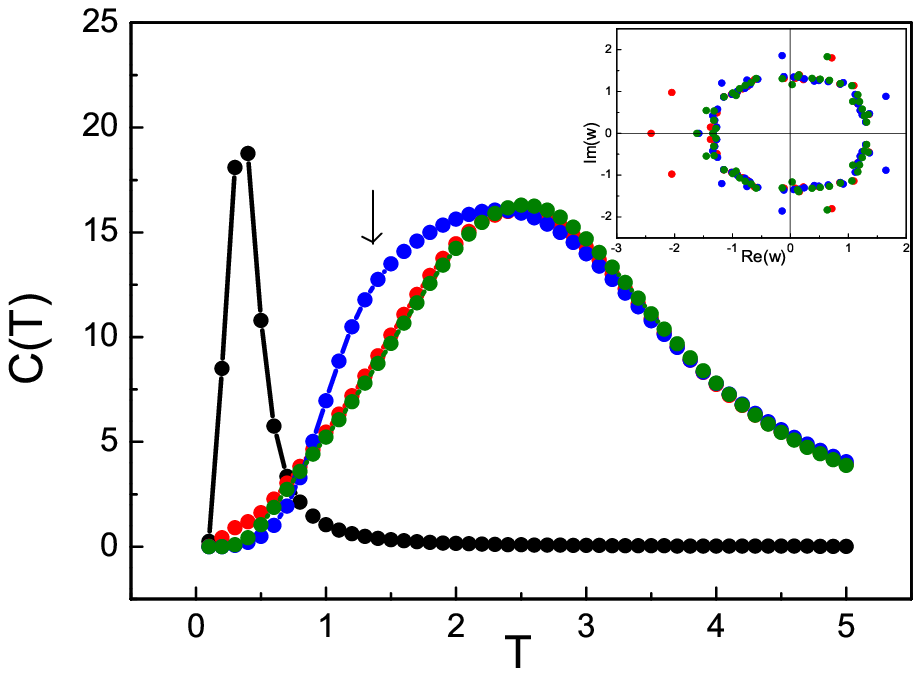}}     
\caption{Specific Heat as a function of temperature. Insets show the zeros of the partition function 
         Eq.(\ref{eq:PF2}) for the energy schemes $U_2, U_3$ and $U_4$. The sharp peak at $T\sim 0.4$
        (black) corresponds to $U_1$. (a) Sequence $S_1$: A Shoulder 
         is seen for energy schemes $U_2$ (red) and $U_4$ (green) at $T\sim 1.0$. (b) Sequence $S_2$: 
         A shoulder is seen for energy schemes $U_2$ (red) and $U_4$ (green) at $T\sim 1.5$; a prominent peak 
         at about the same value of $T$ is seen for energy scheme $U_3$ (blue). 
        (c) Sequence $S_3$: A shoulder is seen at $T\sim 1.4$ only for the 
         scheme $U_3$ (blue).}
         
\label{fig:SpHt}  
\end{figure}
\end{center}

We have computed specific heat as a function of temperature for the three distinct sequences, say $S_1, S_2$ 
and $S_3$, and the energy schemes $U_1 - U_4$; the results are shown 
in Fig.\ref{fig:SpHt}. The insets show the complex zeros, $w_E$, of the partition function defined in 
Eq.(\ref{eq:PF2}). 

As can be seen from the figure, all the three sequences show a sharp peak at $T\sim 0.4$ for the energy scheme $U_1$.
They also show more or less the same behavior for the energy schemes $U_2$ and $U_4$ implying thereby that
these two schemes, except for a scale-shift, are probably the same. Sequence $S_1$ has a hint of a shoulder at
$T\sim 1.0$ (Fig.\ref{fig:SpHt}(a)). On the other hand, sequence $S_2$ shows a prominent peak at $T\sim 1.5$ for
the energy scheme $U_3$ (Fig.\ref{fig:SpHt}(b)) whereas only a shoulder at the same value of $T$ for $U_2$
and $U_4$. Sequence $S_3$ shows a shoulder at $T\sim 1.4$ only for the scheme $U_3$.

The broad peak seen at $T\sim 2.3$ is indicative of a coil-to-globule transition whereas the ones observed at
$T\sim 1.0-1.5$ are indicative of a globule-to-native folding transition. According to Socci and Onuchic \cite{Socci},
fast folding sequences give rise to two different peaks in the specific heat of which the one at lower temperature 
is sharp and prominent. This observation implies that the sequence $S_2$ with the energy scheme $U_3$ is a fast
folder as compared to the others.

It must be mentioned here that configurations with less number of contacts for these sequences and the energy
schemes may have lower energy than the ground state values reported in TABLE II. We summarize the results in
TABLE \ref{tab:steps25IV}. From these tables, it is clear that the designability ({\it i.e}., number of 
designing sequences) 
of these lowest energy, less compact (only 10 contacts) configurations is much less than that of the most 
compact (12 contacts) configurations. 

\begin{table}[ht] 
\begin{tabular}{cccccc}\hline 
$S$ \ & $E_g(U_1)$ \ & $E_g(U_2)$ \ & $E_g(U_3)$ \ & $E_g(U_4)$ \ & CC \\  \hline 
$S_1$[2] \ & -9(4) \ & -63(4) \ & -64(4) \ & -62(66) \ & 1231212313124231432421321 \\
$S_2$[13],$S_3$[25] \ & -9(33) \ & -66(9) \ & -66(9) \ & -66(9) \ & 1231242343421231212432421 \\ \hline
\end{tabular}\\ 
\caption{Diamond lattice walks. Number of contacts = 10. Numbers within square brackets denote the number of 
         distinct configurations;
         those within curved brackets denote the number of designing sequences.
         $E_g(U)$: minimum energy for the scheme $U$; CC: common configuration}
\label{tab:steps25IV}         
\end{table}
A question arises whether shorter chains also have a unique target configuration such as the one shown 
in Fig.\ref{fig:DesConf}; so, we have repeated these calculations for walks consisting of 23, 21 and 17 steps
and have presented the data in TABLE \ref{tab:steps}. We note that the number of minimum-energy configurations common to all the energy-schemes $U_1-U_4$ ({\i.e.,} the ground state degeneracy) increases for shorter walks. 

\begin{figure} [h]
\centering
\unitlength1cm
  \begin{minipage}[t]{6.0 cm}
    \subfigure{\hspace{-1.0 cm} \includegraphics[width=1.5\textwidth ]{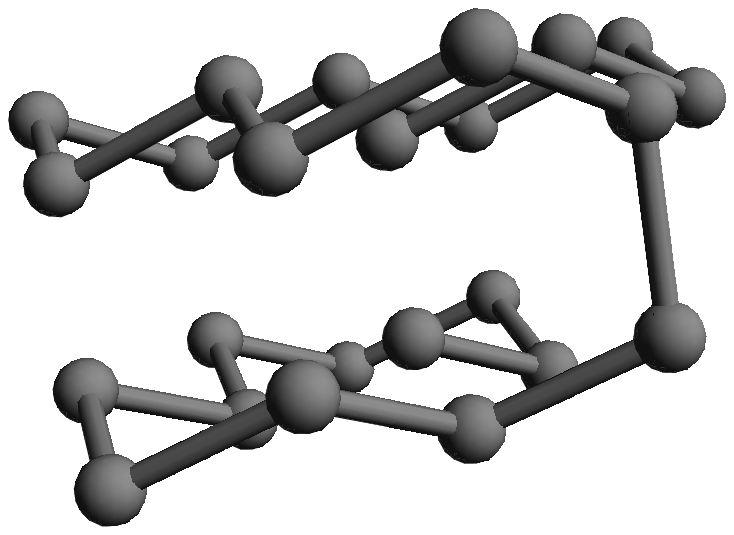}}               
  \end{minipage}
\hfill
  \begin{minipage}[t]{6.0 cm}            
    \subfigure{\hspace{-2.0 cm}  \includegraphics[width=1.5\textwidth ]{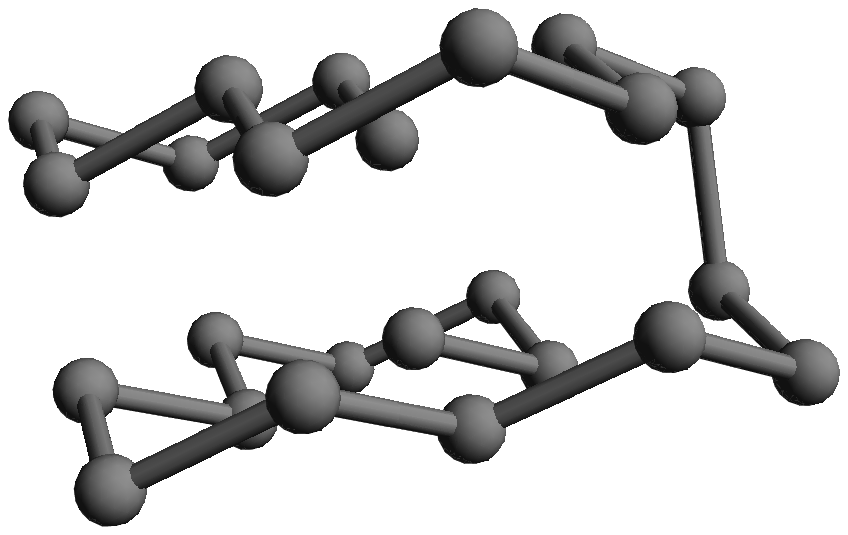}}            
  \end{minipage}
  \vspace{-1.5 cm}         
\caption{23-step configurations ${\cal C}_{23}^{g(1)}$ and ${\cal C}_{23}^{g(2)}$ that are common 
         to all the four energy schemes considered (Eq.(\ref{eq:EgyScheme1}-\ref{eq:EgyScheme4})) . 
         As is evident from these figures, they consist consists of two 'sheets'.}
\label{fig:DesConf23}  
\end{figure}

23-{\it step walk}: There are two 23-step target configurations (TABLE \ref{tab:steps23I}) for which we carried out 
the sequence-design process and the
results are presented in TABLES \ref{tab:steps23II} - \ref{tab:steps23III}. There are two distinct sequences designing 
the configurations, ${\cal C}_{23}^{g(1)}$ and ${\cal C}_{23}^{g(2)}$ - one for the energy-schemes $U_1 \ \& \ U_3$
and the other for the energy-schemes $U_2 \ \& \ U_4$. It is interesting to note that the sequences
that design ${\cal C}_{23}^{g(2)}$ are the {\it inverse} of those designing ${\cal C}_{23}^{g(1)}$; for example,
the sequence $\tilde{S}_1$ designing ${\cal C}_{23}^{g(2)}$ for the energy-schemes
$U_1 \ \& \ U_3$ is the inverse ({\it i.e}., mirror-image) of the sequence $S_1$
designing ${\cal C}_{23}^{g(1)}$ for the same energy-schemes. Even their respective gap-parameters have roughly
the same value. Moreover, the configurations ${\cal C}_{23}^{g(1)}$ and ${\cal C}_{23}^{g(2)}$ shown in 
Fig.\ref{fig:DesConf23} can be mapped into each other as follows.

The reversed configuration code of  ${\cal C}_{23}^{g(2)}$ is 23123231214213132121321; a clockwise permutation
of the direction-vector labels (123) $\to$ (231) {\it and} relabeling them as (123) leads to the code 
12312123134132321313213 which is exactly that of ${\cal C}_{23}^{g(1)}$. Hence, we have effectively a unique
target structure.

\begin{figure} [h]
\centering
\unitlength1cm
  \includegraphics[width=1.0\textwidth ]{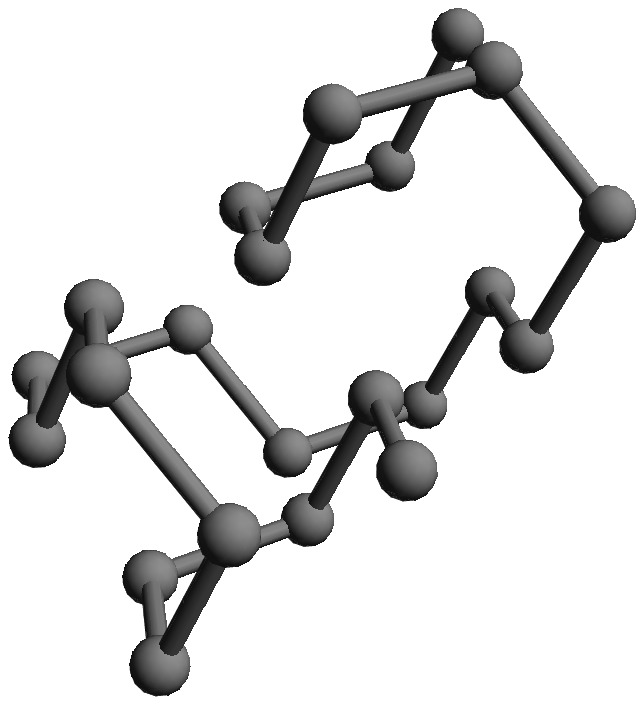} 
  \vspace{-2.0 cm}                  
  \caption{21-step configuration ${\cal C}_{21}^{g(5)}$ that is common 
         to all the four energy schemes considered (Eq.(\ref{eq:EgyScheme1}-\ref{eq:EgyScheme4})) and has
         the highest set of $f$ values. Two 'sheet'-like structures can be seen.}
\label{fig:DesConf21}  
\end{figure}

21-{\it step walk}: There are fifteen 21-step configurations common to all the energy-schemes 
for which we carried out the sequence-design process and the results are presented in TABLE \ref{tab:steps21}. 
We note that
the same squence, $H_2PH_3PH_2P_4H_2P_3(HP)_2$, designs the five configurations ${\cal C}_{21}^{g(3)}$,
${\cal C}_{21}^{g(4)}$, ${\cal C}_{21}^{g(5)}$, ${\cal C}_{21}^{g(6)}$ and ${\cal C}_{21}^{g(9)}$ for the
energy-schemes $U_2-U_4$; the gap-parameter $f$ is also the same. The sequence designing ${\cal C}_{21}^{g(5)}$
for the energy-scheme $U_1$ has the highest value of $f$ ($= 6.4846$). Among the fifteen structures,
${\cal C}_{21}^{g(5)}$ has the highest set of $f$ values and hence may be chosen as the target structure
(Fig.\ref{fig:DesConf21}).

\begin{figure} [h]
\centering
\unitlength1cm
  \includegraphics[width=1.0\textwidth, angle=0]{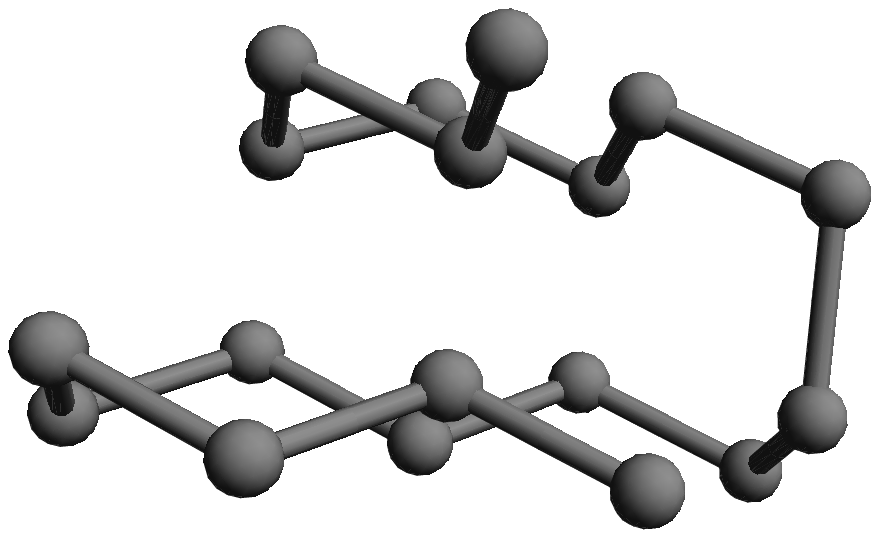}
  \vspace{-3.0 cm}                  
  \caption{17-step configuration ${\cal C}_{12}^{g(5)}$ that is common 
         to all the four energy schemes considered (Eq.(\ref{eq:EgyScheme1}-\ref{eq:EgyScheme4})) and has
         the highest set of $f$ values. Again, 'sheet'-like structures can be seen.}      
\label{fig:DesConf17}  
\end{figure}

17-{\it step walk}: There are sixty eight 17-step configurations common to all the energy-schemes 
for which we carried out the above-described sequence-design process. We found two configurations being
designed by the same sequence ($PHP_3H_3(PH)_2(HP)_3$, for all the energy-schemes) with maximum values for
the gap parameter - data presented in TABLE \ref{tab:steps17}. Further, we note that these configurations 
are mappable into
each other by the the relabeling of the direction vectors $3 \longleftrightarrow 4$ and hence may be 
considered equivalent (Fig.\ref{fig:DesConf17}).

To sum up, in the case when we have ground-state degeneracy, we can choose a target structure by adopting 
either of the two methods or both - namely, (i) checking whether there are configurations whose designing
sequences are inverse of each other and whether these configurations can be mapped into each other by a 
permutation and relabeling of the direction-vectors, and (ii) choose the configurations that have highest 
set of $f$ values. This way, the ground-state degeneracy can be minimized implying thereby an efficient approach
towards realizing a funnel-shaped energy-landscape.   

\begin{table}[t]
\begin{tabular}{ccc}\hline 
N \ & M(N) \ & $Z_N^g$ \\ \hline
25 \ & 12 \ & 1 \\
23 \ & 10 \ & 2 \\
21 \ & 9 \ & 15 \\
17 \ & 6 \ & 68 \\ \hline
\end{tabular}\\
\caption{$Z_N^g$: the number of maximally compact N-step minimum-energy configurations on a diamond lattice,
which are common to all he energy-schemes, $U_1-U_4$. M(N) is the maximum number of contacts for walks of 
length N. }
\label{tab:steps}
\end{table}
\begin{table}[t]
\begin{tabular}[t]{cccc}\hline 
$U$ \ & $Z_{23}^g(U)$ \ & $E_g$ \ & $S_g$\\ \hline 
$U_1$ \ & 34 \ & -8 \ & 10 \\
$U_2$ \ & 34 \ & -56 \ & 10 \\
$U_3$ \ & 34 \ & -58 \ & 10 \\
$U_4$ \ & 2 \ & -54 \ & 1001 \\ \hline 
\end{tabular}\\ 
\caption{ Data for 23-step walks on a diamond lattice; definitions are the same as the ones in 
TABLE \ref{tab:steps25II}. 
The two configurations common to all the energy-schemes are, (i) ${\cal C}_{23}^{g(1)}$: 12312123134132321313213 
and (ii) ${\cal C}_{23}^{g(2)}$: 12312123131241213232132. }
\label{tab:steps23I}
\end{table}
\begin{table}[t]
\begin{tabular}{ccc}\hline 
$U$ \ & Sequence \ & f\\  \hline 
$U_1$ \ & $S_1 \ : \ HP_4(H_2P)_2HPH_3P_2HP_3H_2$ \ & 4.1783\\
$U_2$ \ & $S_2 \ : \ HPHP_2H_2P_2HPH(P_2H_2)_3$ \ & 3.5876\\
$U_3$ \ & $S_1 \ : \ HP_4(H_2P)_2HPH_3P_2HP_3H_2$ \ & 3.8694\\
$U_4$ \ & $S_2 \ : \ HPHP_2H_2P_2HPH(P_2H_2)_3$ \ & 3.2946\\ \hline
\end{tabular}\\ 
\caption{${\cal C}_{23}^{g(1)}$: 12312123134132321313213; definitions are the same as the ones in 
TABLE \ref{tab:steps25III}.}
\label{tab:steps23II}
\bigskip
\begin{tabular}{ccc}\hline 
$U$ \ & Sequence \ & f\\  \hline 
$U_1$ \ & $\tilde{S}_1 \ : \ H_2P_3HP_2H_3PH(PH_2)_2P_4H$ \ & 4.1776\\
$U_2$ \ & $\tilde{S}_2 \ : \ (H_2P_2)_3HPHP_2H_2P_2HPH$ \ & 3.5860\\
$U_3$ \ & $\tilde{S}_1 \ : \ H_2P_3HP_2H_3PH(PH_2)_2P_4H$ \ & 3.8759\\
$U_4$ \ & $\tilde{S}_2 \ : \ (H_2P_2)_3HPHP_2H_2P_2HPH$ \ & 3.2959\\ \hline
\end{tabular}\\ 
\caption{${\cal C}_{23}^{g(2)}$: 12312123131241213232132; definitions are the same as the ones in 
TABLE \ref{tab:steps25III}.}
\label{tab:steps23III}
\end{table}
\begin{table}[ht]
\begin{minipage}[t]{6.5 cm}
\begin{tabular}{c}\hline
${\cal C}_{21}^{g(1)}$: 123413432121341312313 \\ 
\end{tabular}\\
\begin{tabular}{ccc}\hline 
$U$ \ & Sequence \ & f\\  \hline 
$U_1$ \ & $H_2P_4HP_2H_2P_2(HP)_2H_4P$ \ & 5.1075 \\
$U_2$ \ & $(PH)_2HPHP_2H_2P_4HP_2H_4$ \ & 4.3989 \\
$U_3$ \ & $(PH)_2HPHP_2H_2P_4HP_2H_4$ \ & 4.7583 \\
$U_4$ \ & $PH_2P_2H_3PH_2P_4HP_2H_3P$ \ & 3.8996 \\ \hline
\end{tabular}\\ 
\begin{tabular}{c}
${\cal C}_{21}^{g(2)}$: 123423132131431212343 \\ 
\end{tabular}\\
\begin{tabular}{ccc}\hline 
%$U$ \ & Sequence \ & f\\  \hline 
$U_1$ \ & $H_2P_4H_3P(PH)_2P_2H_3PHP$ \ & 4.7633 \\
$U_2$ \ & $H_2P_4H_3P_2HPH_2PH_2P_2HP$ \ & 4.2390 \\
$U_3$ \ & $H_2P_4H_3P_2HPH_2PH_2P_2HP$ \ & 4.9193 \\
$U_4$ \ & $H_4P_2HP_3H(HP)_2H_2P_2HP_2$ \ & 4.7140 \\ \hline
\end{tabular}\\ 
\begin{tabular}{c}
${\cal C}_{21}^{g(3)}$: 123124121343212412312 \\ 
\end{tabular}\\
\begin{tabular}{ccc}\hline 
%$U$ \ & Sequence \ & f\\  \hline 
$U_1$ \ & $(PH)_2P_3H_2PHP_2H_3P_2H_3P$ \ & 6.1679 \\
$U_2$ \ & $H_2PH_3PH_2P_4H_2P_3(HP)_2$ \ & 6.0343 \\
$U_3$ \ & $H_2PH_3PH_2P_4H_2P_3(HP)_2$ \ & 6.5275 \\
$U_4$ \ & $H_2PH_3PH_2P_4H_2P_3(HP)_2$ \ & 5.2154 \\ \hline
\end{tabular}\\ 
\begin{tabular}{c}
${\cal C}_{21}^{g(4)}$: 123124121343212432132 \\ 
\end{tabular}\\
\begin{tabular}{ccc}\hline 
%$U$ \ & Sequence \ & f\\  \hline 
$U_1$ \ & $H_3P_2H_3P_2H_2P_2HP_3H_2P_2$ \ & 6.3509 \\
$U_2$ \ & $H_2PH_3PH_2P_4H_2P_3(HP)_2$ \ & 6.0343 \\
$U_3$ \ & $H_2PH_3PH_2P_4H_2P_3(HP)_2$ \ & 6.5275 \\
$U_4$ \ & $H_2PH_3PH_2P_4H_2P_3(HP)_2$ \ & 5.2154 \\ \hline
\end{tabular}\\ 
\begin{tabular}{c}
${\cal C}_{21}^{g(5)}$: 123124131242313413213 \\ 
\end{tabular}\\
\begin{tabular}{ccc}\hline 
%$U$ \ & Sequence \ & f\\  \hline 
$U_1$ \ & $P_2H_2P_3HP_2H_2(P_2H_3)_2$ \ & 6.4846 \\
$U_2$ \ & $H_2PH_3PH_2P_4H_2P_3(HP)_2$ \ & 6.0343 \\
$U_3$ \ & $H_2PH_3PH_2P_4H_2P_3(HP)_2$ \ & 6.5275 \\
$U_4$ \ & $H_2PH_3PH_2P_4H_2P_3(HP)_2$ \ & 5.2154 \\ \hline
\end{tabular}\\ 
\begin{tabular}{c}
${\cal C}_{21}^{g(6)}$: 123124131242313423123 \\ 
\end{tabular}\\
\begin{tabular}{ccc}\hline 
%$U$ \ & Sequence \ & f\\  \hline 
$U_1$ \ & $(PH)_2P_3H_2PHP_2H_3P_2H_3P$ \ & 6.1679 \\
$U_2$ \ & $H_2PH_3PH_2P_4H_2P_3(HP)_2$ \ & 6.0343 \\
$U_3$ \ & $H_2PH_3PH_2P_4H_2P_3(HP)_2$ \ & 6.5275 \\
$U_4$ \ & $H_2PH_3PH_2P_4H_2P_3(HP)_2$ \ & 5.2154 \\ \hline
\end{tabular}\\
\begin{tabular}{c}
${\cal C}_{21}^{g(7)}$: 123243132131231412323 \\ 
\end{tabular}\\
\begin{tabular}{ccc}\hline 
%$U$ \ & Sequence \ & f\\  \hline 
$U_1$ \ & $P_2HP_3H_2PH_3P_2H_2P_2H_3P$ \ & 4.2178 \\
$U_2$ \ & $P_2HP_3H_2PH_3P_2H_2P_2H_3P$ \ & 4.1126 \\
$U_3$ \ & $P_2HP_3H_2PH_3P_2H_2P_2H_3P$ \ & 4.3714 \\
$U_4$ \ & $H_4P_2HP_3H_2(PH)_2(HP_2)_2$ \ & 4.7140 \\ \hline
\end{tabular}\\
\begin{tabular}{c}
${\cal C}_{21}^{g(8)}$: 123213232141321312313 \\ 
\end{tabular}\\
\begin{tabular}{ccc}\hline 
%$U$ \ & Sequence \ & f\\  \hline 
$U_1$ \ & $(HP)_2PHPH_2P_2(HP)_2H_3P_2HP$ \ & 5.6569 \\
$U_2$ \ & $(HP)_2PHPH_2P_2(HP)_2H_3P_2HP$ \ & 4.7329 \\
$U_3$ \ & $(HP)_2PHPH_2P_2(HP)_2H_3P_2HP$ \ & 5.1549 \\
$U_4$ \ & $HPH_3P_2H_2P_3(PH)_3P_2H_2$ \ & 3.8735 \\ \hline
\end{tabular}\\
\end{minipage}
\hfill
\begin{minipage}[t]{6.5 cm}
\begin{tabular}{c}\hline
${\cal C}_{21}^{g(9)}$: 124123121434212342142 \\ 
\end{tabular}\\
\begin{tabular}{ccc}\hline 
$U$ \ & Sequence \ & f\\  \hline 
$U_1$ \ & $H_3P_2H(H_2P_2)_2HP_3H_2P_2$ \ & 6.3509 \\
$U_2$ \ & $H_2PH_3PH_2P_4H_2P_3(HP)_2$ \ & 6.0343 \\
$U_3$ \ & $H_2PH_3PH_2P_4H_2P_3(HP)_2$ \ & 6.5275 \\
$U_4$ \ & $H_2PH_3PH_2P_4H_2P_3(HP)_2$ \ & 5.2154 \\ \hline
\end{tabular}\\ 
\begin{tabular}{c}
${\cal C}_{21}^{g(10)}$: 121234314321241312313 \\ 
\end{tabular}\\
\begin{tabular}{ccc}\hline 
%$U$ \ & Sequence \ & f\\  \hline 
$U_1$ \ & $HPH_3P_3HP(PH)_3P_2HPH_2$ \ & 5.2511 \\
$U_2$ \ & $P_2H_2P(PH)_2P_2H_2P_2HP_2H_4$ \ & 4.0692 \\
$U_3$ \ & $HPH_3P_3HP(PH)_3P_2HPH_2$ \ & 4.9029 \\
$U_4$ \ & $PH_3P_2HP(PH)_2P_3H_2PH_3P$ \ & 3.7262 \\ \hline
\end{tabular}\\
\begin{tabular}{c}
${\cal C}_{21}^{g(11)}$: 121343413431231413414 \\ 
\end{tabular}\\
\begin{tabular}{ccc}\hline 
%$U$ \ & Sequence \ & f\\  \hline 
$U_1$ \ & $H_2P_2H_2P_4HP_2HPH_3(HP)_2$ \ & 4.4721 \\
$U_2$ \ & $H_2PH_3P_4HP_2HPH_3P_2HP$ \ & 3.9009 \\
$U_3$ \ & $H_2PH_3P_4HP_2HPH_3P_2HP$ \ & 4.2762 \\
$U_4$ \ & $P_2HP_2H(HP)_2H_2P_3HP_2H_4$ \ & 4.0415 \\ \hline
\end{tabular}\\
\begin{tabular}{c}
${\cal C}_{21}^{g(12)}$: 121321234231321313242 \\ 
\end{tabular}\\
\begin{tabular}{ccc}\hline 
%$U$ \ & Sequence \ & f\\  \hline 
$U_1$ \ & $H_2P_4(HP)_2H_2PH(HP)_2(PH)_2$ \ & 3.6181 \\
$U_2$ \ & $PHP_2H_5P_2HP_2(P_2H_2)_2$ \ & 3.6856 \\
$U_3$ \ & $H(HP)_2H_2P_3H_2PHP_2HP_2H_2P$ \ & 4.6512 \\
$U_4$ \ & $H_4P_2HP_3H_2(PH)_2(HP_2)_2$ \ & 4.7140 \\ \hline
\end{tabular}\\
\begin{tabular}{c}
${\cal C}_{21}^{g(13)}$: 121321231242313123132 \\ 
\end{tabular}\\
\begin{tabular}{ccc}\hline 
%$U$ \ & Sequence \ & f\\  \hline 
$U_1$ \ & $PHP_2H_3(PH)_2P_2H(HP)_2(PH)_2$ \ & 4.7329 \\
$U_2$ \ & $PHP_2H_5PHP_2H_2P_3(PH)_2$ \ & 4.0565 \\
$U_3$ \ & $PHP_2H_5PHP_2H_2P_3(PH)_2$ \ & 4.4721 \\
$U_4$ \ & $PHP_2H_3PH_3P_2H_2P_3(PH)_2$ \ & 3.5820 \\ \hline
\end{tabular}\\
\begin{tabular}{c}
${\cal C}_{21}^{g(14)}$: 121321241232314124132 \\ 
\end{tabular}\\
\begin{tabular}{ccc}\hline 
%$U$ \ & Sequence \ & f\\  \hline 
$U_1$ \ & $PH_3P_2HP_4H_5P_3H_2P$ \ & 4.5210 \\
$U_2$ \ & $PH_3(P_2H)_2PH_2PH_2P_3H_2P$ \ & 4.2083 \\
$U_3$ \ & $PH_3P_2HP_4H_5P_3H_2P$ \ & 4.4488 \\
$U_4$ \ & $H_2PH(P_2H)_2PH_2PHP_2HPH_2P$ \ & 4.3894 \\ \hline
\end{tabular}\\
\begin{tabular}{c}
${\cal C}_{21}^{g(15)}$: 121321243231421413232 \\ 
\end{tabular}\\
\begin{tabular}{ccc}\hline 
%$U$ \ & Sequence \ & f\\  \hline 
$U_1$ \ & $H_4P_2HP(P_2H)_2(H_2P_2)_2$ \ & 4.5518 \\
$U_2$ \ & $PH_3P_2HPH_3P_2(P_2H)_2HPH$ \ & 4.4783 \\
$U_3$ \ & $H_4P_2HP(P_2H)_2(H_2P_2)_2$ \ & 5.2152 \\
$U_4$ \ & $H_3(HP_2)_2(PH)_2HPHP_2H_2P_2$ \ & 3.8307 \\ \hline
\end{tabular}\\
\end{minipage}
\caption{${\cal C}_{21}^g$: 21-step ground state configurations common to all the energy-schemes.
         definitions are the same as the ones in TABLE \ref{tab:steps25III}}
\label{tab:steps21}         
\end{table}
\begin{table}[t]
\begin{tabular}{ccc}\hline 
$U$ \ & Sequence \ & f\\  \hline 
$U_1$ \ & $PHP_3H_3(PH)_2(HP)_3$ \ & 3.9903\\
$U_2$ \ & $PHP_3H_3(PH)_2(HP)_3$ \ & 4.2740\\
$U_3$ \ & $PHP_3H_3(PH)_2(HP)_3$ \ & 4.4030\\
$U_4$ \ & $PHP_3H_3(PH)_2(HP)_3$ \ & 4.0885\\ \hline
\end{tabular}\\ 
\caption{Designing sequence for 17-step configurations, ${\cal C}_{17}^{g(1)}$: 12132121341312313 and 
${\cal C}_{17}^{g(2)}$: 12142121431412414. f: maximum gap-parameter. Subscript of H or P denotes the number 
of times the corresponding symbol is repeated.}
\label{tab:steps17}
\end{table}

\bigskip

\noindent 3. {\it Square Lattice walks}:

\bigskip

Just to check whether the algorithm described above for diamond lattice walks also works for a square lattice
(same coordination number), we have considered 24-step walks on a square lattice.

There are a total of 1081 24-step maximally compact (16 contacts) SAWs on a square lattice with distinct
contact-maps. By fitting all possible 25-bit binary sequences (12 ones and 13 zeros) to each of these configurations,
we found one configuration, 112234323341434412141123 \cite{footnote5}, that is the minimum energy configuration
for all the four energy schemes $U_1 - U_4$.

\begin{figure} [ht]
\centering
\unitlength1cm
  \begin{minipage}[t]{5.75 cm}
    \subfigure [] {\includegraphics[width=1.0\textwidth ]{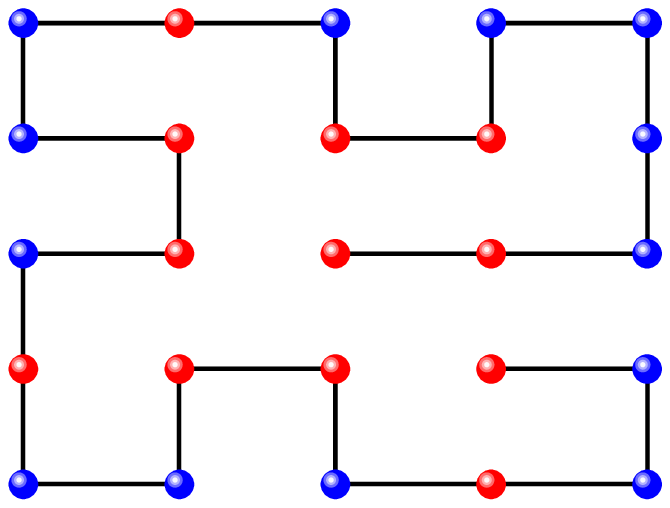}}                  
  \end{minipage}
\hfill
  \begin{minipage}[t]{5.75 cm}            
    \subfigure [] {\includegraphics[width=1.1\textwidth ]{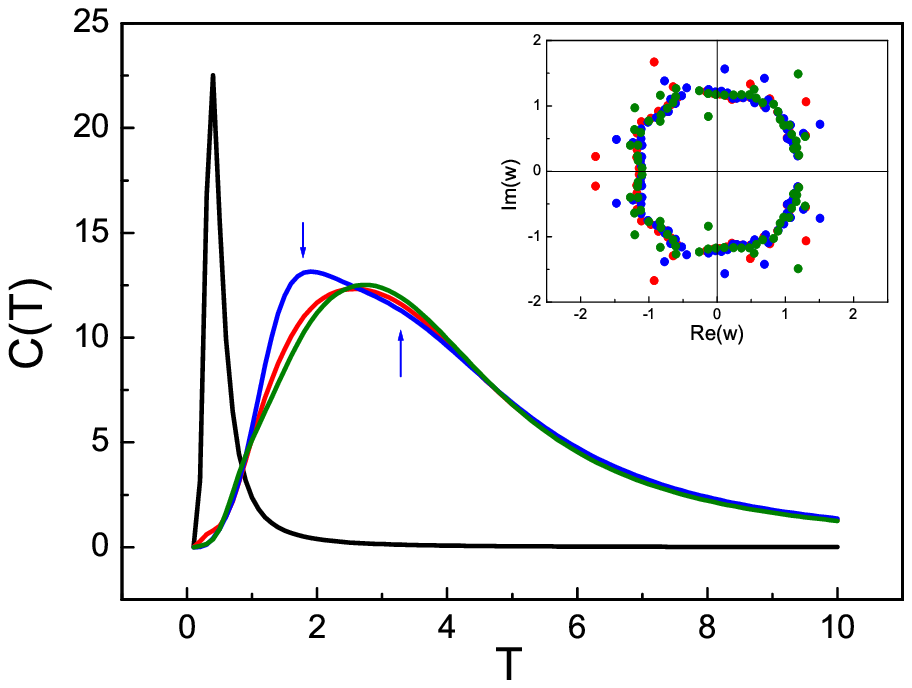}}               
  \end{minipage}
\caption{(a) The ground state configuration 112234323341434412141123 displaying the designed sequence,
         $H_2P_4H_2PHP_2H_2PHP_2H_2PHP_2H$ ($H$: red; $P$: blue).
         (b)  Specific Heat as a function of temperature. Insets show the zeros of the partition function 
         Eq.(\ref{eq:PF2}) for the energy schemes $U_2, U_3$ and $U_4$. The sharp peak at $T\sim 0.4$
        (black) corresponds to $U_1$. A single broad peak at $T\sim 2.5$ is 
         seen for both $U_2$ (red) and $U_4$ (green). In contrast, data for $U_3$ (blue) show a prominent peak at 
         $T\sim 1.5$ besides a broad peak at $T\sim 3$ (indicated by arrows).}      
\label{fig:SpHtSq1}  
\end{figure}

For all these energy schemes, we found just 4 sequences common to the sets of sequences designing this configuration.
We estimated the gap-parameters for these sequences by using the whole set of maximally compact walks, and found
that just one of these four sequences has the maximum gap-parameter, $f$, for $U_1 - U_4$ (TABLE \ref{tab:SqTab}). 
The corresponding ground state configuration is shown in Fig.\ref{fig:SpHtSq1}(a).

Using the DoS, computed for this designed sequence and energy schemes, we have calculated the specific heat and
presented the data in Fig.\ref{fig:SpHtSq1}(b). The prominent first peak seen for $U_3$ implies that 
this designed sequence is a fast folder only for this energy scheme.

There are six distinct less compact configurations with 14 contacts having less energy ($= -78$) for this designed 
sequence than the ground state energy ($= -76$) of the maximally compact walks, but only for the energy scheme $U_4$.
While the number of sequences designing the maximally compact ground state configuration (Fig.\ref{fig:SpHtSq1}(a))
is 220, it is much less ($=32$) for these less compact walks. In other words, the latter are not only less designable
for $U_4$ but are actually excited states for the other energy schemes.

\begin{figure} [ht]
\centering
\unitlength1cm
  \begin{minipage}[t]{5.75 cm}
    \subfigure [] {\includegraphics[width=1.0\textwidth ]{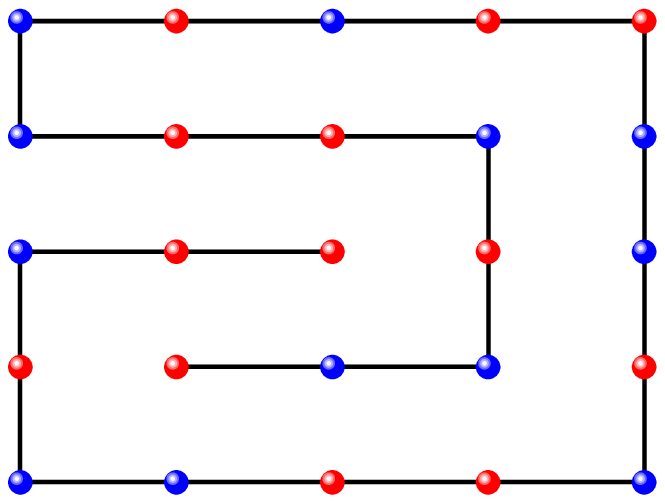}}                  
  \end{minipage}
\hfill
  \begin{minipage}[t]{5.75 cm}            
    \subfigure [] {\includegraphics[width=1.0\textwidth ]{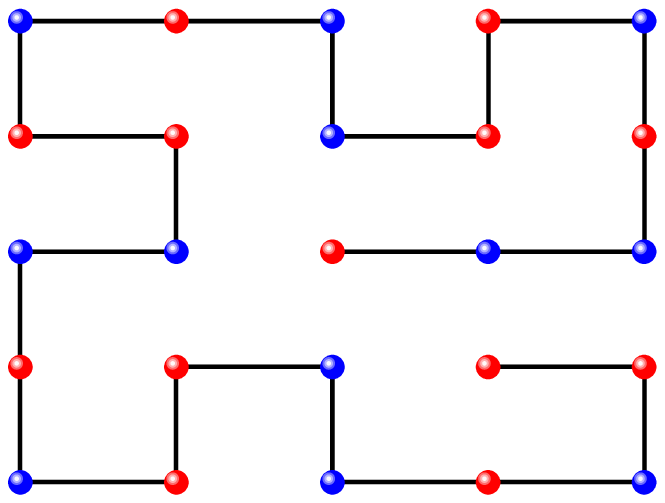}}               
  \end{minipage}
\caption{The designed sequence is $HP_2HPH_2P_2HPH_2P_2HPH_2P_2HPH_2$ ($H$: red; $P$: blue).
         (a) The ground state configuration 112233321111444433332211 displaying the designed sequence,
         is only for the scheme $U_4$ with a designability of 195 sequences. (b) The ground state 
         configuration 112234323341434412141123 for all the energy schemes. While this sequence is the 
         only sequence identifying this configuration as the ground state for schemes $U_1 - U_3$, it
         turned out to have the maximum gap-parameter ($= 4.4151$) for the scheme $U_4$.}          
\label{fig:SpHtSq2}  
\end{figure}

We also found that the designability also depends on the number of ones ($H$ type) in the sequence. For example,
ground state configurations for a designed sequence consisting of 13 ones are shown in Fig.\ref{fig:SpHtSq2}(a,b).
Both are designed by the same sequence $HP_2HPH_2P_2HPH_2P_2HPH_2P_2HPH_2$.
The ground state displayed in Fig.\ref{fig:SpHtSq2}(a) is only for the scheme $U_4$ with a designability of 
195 sequences, whereas the one displayed in Fig.\ref{fig:SpHtSq2}(b) is common to all the energy schemes. But
the designability of the latter is just one sequence for the schemes $U_1 - U_3$. We also see that these
configurations don't have a clear hydrophobic interior as compared to the one, Fig.\ref{fig:SpHtSq1}(a), designed 
with 12 H-types.

\begin{table}
\begin{tabular}{ccc}\hline 
U \ & $E_g(U)$ \ & f \\  \hline 
$U_1$ \ & -11 \ & 5.1636 \\
$U_2$ \ & -80 \ & 4.8322 \\ 
$U_3$ \ & -84 \ & 5.1958 \\
$U_2$ \ & -76 \ & 4.3359 \\ \hline
\end{tabular}\\ 
\caption{Square lattice walks. Number of contacts = 16. Common designing sequence 
is $H_2P_4H_2PHP_2H_2PHP_2H_2PHP_2H$; designable ground state configuration:
112234323341434412141123. $E_g(U)$: minimum energy for the scheme $U$; f: maximum value of the gap-parameter.}
\label{tab:SqTab}  
\end{table}

\section*{\leftline{\gross Conclusions}}

We have presented an exact enumeration study of how the sequence-design of maximally compact lattice SAWs 
is influenced by the various contact-energy schemes chosen; the motivation is to evolve an algorithmic 
criterion for identifying the best scheme from among the set of schemes being considered. Moreover, since
the various contact-energy schemes considered in the literature are based on statistical estimates of the average
effective inter-residue interactions of real proteins, it is of interest to know if the target configuration
and its designed sequence is stable with respect to changes in the contact-energy scheme.
In fact, the many-to-one mapping of amino acid sequences onto the set of 
all native structures implies that a given native structure should not be very sensitive to the set of
contact-energy schemes underlying the set of sequences designing it.

We have considered four different schemes, $U_1 - U_4$ (Eq.(\ref{eq:EgyScheme1}-\ref{eq:EgyScheme4}), of 
which the last one has been deliberately chosen not to satisfy the inequality  $(u_{HH} + u_{PP}) < 2u_{HP}$
that the first three satisfy. The enumeration data for maximally compact 25-step walks on a diamond lattice, 
as well as for 24-step walks on a square lattice, suggest the following.

(a) The target structure ({\it i.e}., the most designable, maximally compact configuration) is the same for all
the energy schemes considered: (i) On a diamond lattice, such a configuration consists of two sheet-like secondary   
structures as is evident from Fig.\ref{fig:DesConf}; (ii) On a square lattice also, it is independent of the 
energy scheme and is shown in Fig.\ref{fig:SpHtSq1}(a), which is better designable with 12 rather
than with 13 H-type monomers Fig.\ref{fig:SpHtSq2}(a). {\it i.e}.,designability of the target structure, rather
than the bare structure itself, could very well depend on the proportion of H-type monomers in the sequence. 

(b) The designing sequence may depend on the contact-energy scheme: (i) On a diamond lattice, we find that the
designed sequences for $U_2$ and $U_4$ are the same, implying thereby that these two energy schemes are the  
same from sequence-design point of view, whereas those for $U_1$ and $U_3$ are different (TABLE III); (ii) On a 
square lattice, however, the designed sequence (12 H-type) is the same for all the energy schemes (TABLE V).

(c) We find that the maximally compact ground state configuration is better designable than a few less compact
configurations that are assigned lower energy by the designed sequence.

(d) On the basis of an exact count of the Density of States, we have calculated the temperature-dependent
specific heat for these designed sequences and the four energy schemes. (i) In the case of diamond lattice walks,
sequence $S_2$ with scheme $U_3$ seems to be the best fast-folder because it has a prominent low-temperature
peak in the specific heat that corresponds to the compact-to-native transition; (ii) In the case of square
lattice walks also, the designed sequence (common to all energy schemes) with $U_3$ seems to be the best folder  
for the same reason. These specific heat data imply that the energy scheme $U_3$ is preferable to the other
three.

(e) We have confirmed that this algorithm works for shorter walks as well by presenting data for 23-, 21- and 
17-step walks on a diamond lattice. We have been able to identify a unique target configuration by taking care of 
the ground-state degeneracy as described in the last section. Interestingly, all these shorter target
configurations also show sheet-like secondary structures.

To sum up, we have presented an algorithm for identifying the target configuration that is the minimum energy
configuration for a given set of contact-energy schemes; the process is implicitly sequence-dependent. The best
designing HP-sequence for this configuration is then the one for which the gap-parameter has the maximum value.
That an implicit sequence-dependence of a folding process ensures a realization of a funnel-like energy landscape  
has been demonstrated recently by Chickenji {\it et al} \cite{Chickenji}. In particular, they have shown that 
short fragments excised from proteins suggest sequence-dependent local restrictions in the conformational space 
that are also relatively free from the inaccuracy of the local interactions. While their method assembles 
appropriate short fragments into a target configuration, our method is a sequence-dependent search procedure
for a target configuration that is not strongly dependent on the contact-energy scheme. There is a need for a 
more detailed study of our method {\it vis-a-vis} others. 

\bigskip

SLN thanks Artur Baumgaertner for helpful discussions and a critical reading of the manuscript. He also thanks
Venkatesh Shenoi for bringing the reference \cite{Chickenji} to his attention.

\bigskip

%\newpage


\begin{thebibliography}{99}
\bibitem{footnote2} Our exact enumeration code not only generates all SAWs up to a specified maximum
                    length but also computes the average radius of gyration and the average 
                    end-to-end distance as a function of the length of the walk; it also counts and
                    tabulates, for each length, the number of walks as a function of the number of 
                    contacts made. For a given binary sequence and an energy scheme, it can compute
                    the Density of States too. For a maximum of 25 steps, this program takes roughly 
                    8 hours of a desktop PC's CPU time (Intel Core2 Duo E8400 @ 3 GHz; 4 GB RAM) on
                    Windows XP platform.
\bibitem{footnote3} The actual number of walks is twelve times $Z_N(m)$, because the first two steps
                    have been fixed to be in the directions $v_I(1)$ and $v_{II}(2)$ respectively.
\bibitem{footnote4} The sum is only over the magnitude of $E$ because none of the values of $E$ is a nonzero 
                    positive integer.
\end{thebibliography}
\end{document}